\documentclass[twocolumn,prl,showpacs,amsmath,amssymb]{revtex4}
\usepackage{graphicx}
\usepackage{amssymb,amsthm,amsfonts,amstext}
\usepackage{url}
\def\be{\begin{equation}}
\def\ee{\end{equation}}
\def\bea{\begin{eqnarray}}
\def\eea{\end{eqnarray}}
\def\bma{\begin{mathletters}}
\def\ema{\end{mathletters}}

\def\0{\overline{0}}

\def\q0{\underline{0}}

\def\H{{\cal H}}

\def\S{{\cal S}}

\def\id{{\mathbb I}}

\def\H{{\cal H}}

\def\tr{\mbox{tr}}
\def\one{\leavevmode\hbox{\small1\normalsize\kern-.33em1}}

\def\bra#1{\langle#1|} \def\ket#1{|#1\rangle}

\def\proj#1{\ket{#1}\!\bra{#1}}

\newtheorem{theo}{Theorem}

\newtheorem{prop}[theo]{Proposition}

\def\id{{\mathbb I}}
\def\tr{\mbox{tr}}

\begin{document}

\title{A complete criterion for separability detection}

\author{Miguel Navascu\'es$^1$, 
        Masaki Owari$^{1,2}$ and 
        Martin B. Plenio$^{1,2}$}
\affiliation{$^1$Institute for Mathematical Sciences, Imperial College London, SW7 2PE, UK,\\
$^2$QOLS, The Blackett Laboratory, Imperial College London,
 Prince Consort Rd., SW7 2BW, UK}

\begin{abstract}
Using new results on the separability properties of bosonic systems, we provide a new complete criterion for separability. 
This criterion aims at characterizing the set of separable states from the inside by means of a sequence of efficiently solvable semidefinite programs. 
We apply this method to derive arbitrarily good approximations to the optimal measure-and-prepare strategy in generic state estimation problems. Finally, we report its performance in combination with the criterion developed by Doherty et al. \cite{doherty2} for the calculation of the entanglement robustness of a relevant family of quantum states whose separability properties were unknown.
\end{abstract}
\maketitle       

Research on separability criteria, that is, on computational methods to determine whether a given state is separable or entangled,
is a popular subject in Quantum Information Theory.
Starting from the famous PPT \cite{P96} (Positive Partial Transpose) criterion, 
a considerable number of different separability criteria have been discovered (see the references in \cite{intro_measures,HHHH07,GT08,I07}). Unfortunately, the most efficient ones happen to be \emph{partial} criteria, in the sense that they only detect entanglement or separability in certain situations. This is not surprising, in view of the fact that the separability problem is NP-hard \cite{G03}, \cite{sevag}.

A \emph{complete} criterion for entanglement detection is an algorithm or method that allows to characterize the set of separable states with arbitrary precision. Alternatively, we may say that a complete criterion for separability can solve any instance of the Weak Membership Problem of separability \cite{I07}, i.e., the problem of determining if a given quantum state is close to the core of the set of separable states. Up to now, there exist several different complete separability criteria \cite{witness,HB07,brandao,relaxations,S07,doherty2,doherty}. In all these methods, the set of separable states is approximated successively by an appropriate sequence of sets of states. The complexity of characterizing each of these sets increases as we move along the sequence, that converges to the set of separable states in the asymptotic limit. That way, we can characterize the set of separable states up to a precision that is only limited by our computational resources.

While traditionally Entanglement Theory has focused most of its efforts in developing criteria for entanglement detection \cite{witness,relaxations,doherty2,doherty}, thus characterizing the set of separable states from the \emph{outside}, there exist a few complete criteria that try to approximate the set of separable states from the \emph{inside}, i.e., whose aim is to detect separability instead of entanglement. Examples of the latter type are the algorithms invented by Hulpke et al. \cite{HB07}, Brand\~{a}o et al. \cite{brandao} or Spedalieri \cite{S07}. The main drawbacks of the first two are their high time complexity, as estimated in \cite{I07}. The method designed by Spedalieri, although very promising, does not currently have any proven bounds on its speed of convergence, and cannot be extended to deal with multipartite entanglement \cite{S07}.

Let us also remark that, in all the above cases, the main algorithm works by almost directly invoking the definition of separability, i.e., no insight from Entanglement Theory itself was employed in their conception.

In this paper, we present a new complete criterion for separability which takes inspiration from the symmetric extension criterion developed by Doherty, Parrilo and Spedalieri (the DPS criterion) \cite{doherty,doherty2}. However, while the DPS criterion tries to approximate the set of separable states from the outside, our new criterion will work from the inside.
Our criterion not only detects separability, but also provides an explicit separable decomposition of the separable states in terms of an integral over the Haar measure. Moreover, it has the same proven time complexity as the original DPS criterion. Due to its internal structure, the new criterion can be easily modified to perform linear optimizations over entanglement breaking channels. In particular, given any state estimation problem, the method can output a sequence of measure-and-prepare strategies arbitrarily close to optimal. In the last pages of this letter, as an illustration of its efficiency, we will show its performance by computing the robustness of entanglement of a family of quantum states.


But, first, some remarks about notation. In this letter, we will be mainly concerned with a finite dimensional bipartite Hilbert space $\H \stackrel{\rm def}{=} \H _A \otimes \H_B$.
The set of all linear operators acting on $\H$ will be denoted as ${\cal B}(\H )$, and we will use the term \emph{state} in order to refer to \emph{normalized} non-negative operators in ${\cal B}(\H )$. Finally, the cone of \emph{separable} operators, i.e., the conical combination of all pure product states $\{\proj{\psi_A}\otimes \proj{\psi_B}\}$, will be called ${\cal S}$. Note that any operator in $\S$ must be necessarily non-negative.

The sequences of sets employed to approximate $\S$ in the DPS criterion are either the sets $\{\S^{N}\}$ of $N$-(Bose) symmetrically extendible operators
or the sets $\{\S^{N}_p\}$ of PPT $N$-(Bose) symmetrically extendible operators. 
 These sets are defined as follows: \\
$\Lambda_{AB}\in {\cal B}(\H_{AB} )$ belongs to $\S^{N}$ 
iff there exists an operator $\Lambda_{AB^N}\in B(\H_{AB^N})$ that satisfies the following three conditions:
\begin{enumerate}
\item $\Lambda_{AB^N}\geq 0$.
\item $\tr_{B^{N-1}}(\Lambda_{AB^N})=\Lambda_{AB}$.

\item$\Lambda_{AB^N}$ is Bose symmetric in $\H_B^{\otimes N}$, i.e., $\Lambda_{AB^N}(\id_A\otimes P_{\mbox{sym}}^N)=\Lambda_{AB^N}$, 
where $P_{\mbox{sym}}^N$ denotes the projector onto the symmetric subspace of $\H_B^{\otimes N}$.
\end{enumerate}
Similarly, $\Lambda_{AB} \in \S^{N}_p$ iff there exists an operator $\Lambda_{AB^N}\in B(\H_{AB^N})$ fulfilling 1-3 and the additional constraint:

\begin{enumerate}
\setcounter{enumi}{3}
\item $\Lambda_{AB^N}$ has a Positive Partial Transpose (i.e., it is PPT) \cite{P96} w.r.t. the bipartition $AB^{\lceil N/2\rceil}|B^{\lfloor N/2\rfloor}$.
\end{enumerate}
Doherty et al. \cite{doherty} proved that both sequences $\{ \S^{N} \}_{N=1}^{\infty}$ and $\{ \S^{N}_p \}_{N=1}^{\infty}$
converge to $\S$ from the outside: 
\begin{equation}
\S^1_{(p)}\supset \S^2_{(p)}\supset \S^3_{(p)}\supset... \supset \S, \mbox{with }\lim_{N\to\infty} \S^N_{(p)}=\S.
\label{secuencia}
\end{equation}

The DPS criterion consists, precisely, in checking if $\rho_{AB}\in \S^N_{(p)}$ for all $N$. 
From the relations above, the DPS criterion is clearly complete:
if $\rho_{AB}$ is entangled, 
then there exists an $M$ such that $\rho_{AB}\not\in \S^M_{(p)}$, so all entangled states can be eventually detected.
As all these sets are defined through linear matrix inequalities, 
the problem of determining whether a given state belongs to one of them can be cast as a semidefinite program (SDP) \cite{sdp}.


In \cite{pra} it was shown that, for any $\rho_{AB}\in \S^N$ or $\S^N_p$, 
a small perturbation on system $B$ results in a separable state $\tilde{\rho}_{AB}$. 
More concretely, define $d\equiv \dim(\H_B)$ and
\be
\epsilon_N\equiv\frac{d}{2(d-1)}\min\{1-x:P_{\lfloor N/2\rfloor+1}^{(d-2,N \mod 2)}(x)=0\},
\ee
\noindent with $P^{(\alpha,\beta)}_n(x)$ being the \emph{Jacobi polynomials} \cite{abramo}. Then, the following closed cones of operators
\begin{eqnarray}
& &\hspace*{-0.5cm}\tilde{\S}^N\equiv\{\frac{N}{N+d}\sigma_{AB}+\frac{1}{N+d}\sigma_A\otimes\id_B:\sigma_{AB}\in \S^N\},\label{bosym}\\
& &\hspace*{-0.5cm}\tilde{\S}_p^N\equiv \{(1-\epsilon_N)\sigma_{AB}+\epsilon_N\sigma_A\otimes\frac{\id_B}{d}:\sigma_{AB}\in \S_p^N\},
\label{interior}
\end{eqnarray}
\noindent satisfy $\tilde{\S}^N,\tilde{\S}^N_p\subset \S$, for all $N$. 
Moreover, for any given $\rho_{AB} \in \tilde{\S}_{(p)}^N$, we can find a corresponding separable decomposition in terms of an integral over a Haar measure. Suppose, for instance, that $\sigma_{AB^N}$ is a Bose symmetric extension of $\sigma_{AB}\in \S^N$. Then, the separable decomposition of the state $\rho_{AB} \in \tilde{\S}^N$ that results from perturbing $\sigma_{AB}$ as in equation (\ref{bosym}) is given by
\be
\rho_{AB}= C \int d\phi_B \ \tr _{B^{N}} \left \{ I_A \otimes \phi_B^{\otimes N} \ \sigma_{AB^N} \right \} \otimes \phi_B, \label{eq: separable decomposition} 
\ee
where $C$ is a normalization constant and $d\phi_B$ is a Haar measure over all pure states $\phi_B$ on $\H_B$.
The reader can find a similar decomposition for any state $\rho_{AB} \in \tilde{\S}_{p}^N$ in \cite{pra}.

Note that, in both definitions, taking the limit $N>>d$ implies, 
on one hand, that $\tilde{\S}^N\approx \S^N,\tilde{\S}_p^N\approx \S^N_p$
\footnote{Notice that 
$\epsilon_N\approx \frac{dj^2_{d-2,1}}{(d-1)N^2}$ for $N>>1$ \cite{abramo}, where $j_{d-2,1}$ denotes the first positive zero of the Bessel function $J_{d-2}(x)$. 
It follows that $\lim_{N\to\infty}\epsilon_N=0$.}, and, on the other hand, that $\S^N_p,\S^N\approx \S$. 
It follows that the closures of the limiting sets $\lim_{N\to\infty}\tilde{\S}^N,\tilde{\S}_p^N$ coincide with the set of separable states.

Therefore, as opposed to the DPS criterion, the sequences of sets $\{ \tilde{\S}^N \}_N, 
\{ \tilde{\S}_p^N \}_N$ converge to the set of separable states from the 
\emph{inside}. 
These sets can be characterized using semidefinite programming; 
in fact, a small modification of the computer codes employed to search over the sets $\S^N, \S_p^N$ 
allows to perform optimizations over the sets $\tilde{\S}^N, \tilde{\S}_p^N$. 
We thus arrive at a novel algorithm for entanglement detection complementary to the DPS criterion, 
and with a very similar structure \footnote{In this case, though, we do not have a hierarchy anymore, i.e., in general, 
$\tilde{\S}^N\not\subset\tilde{\S}^{N+1}$.}. 
In view to its strong link with the latter criterion, from now on we will refer to this new algorithm as \emph{DPS$^*$}.

\begin{figure}
  \centering
  \includegraphics[width=8.5 cm]{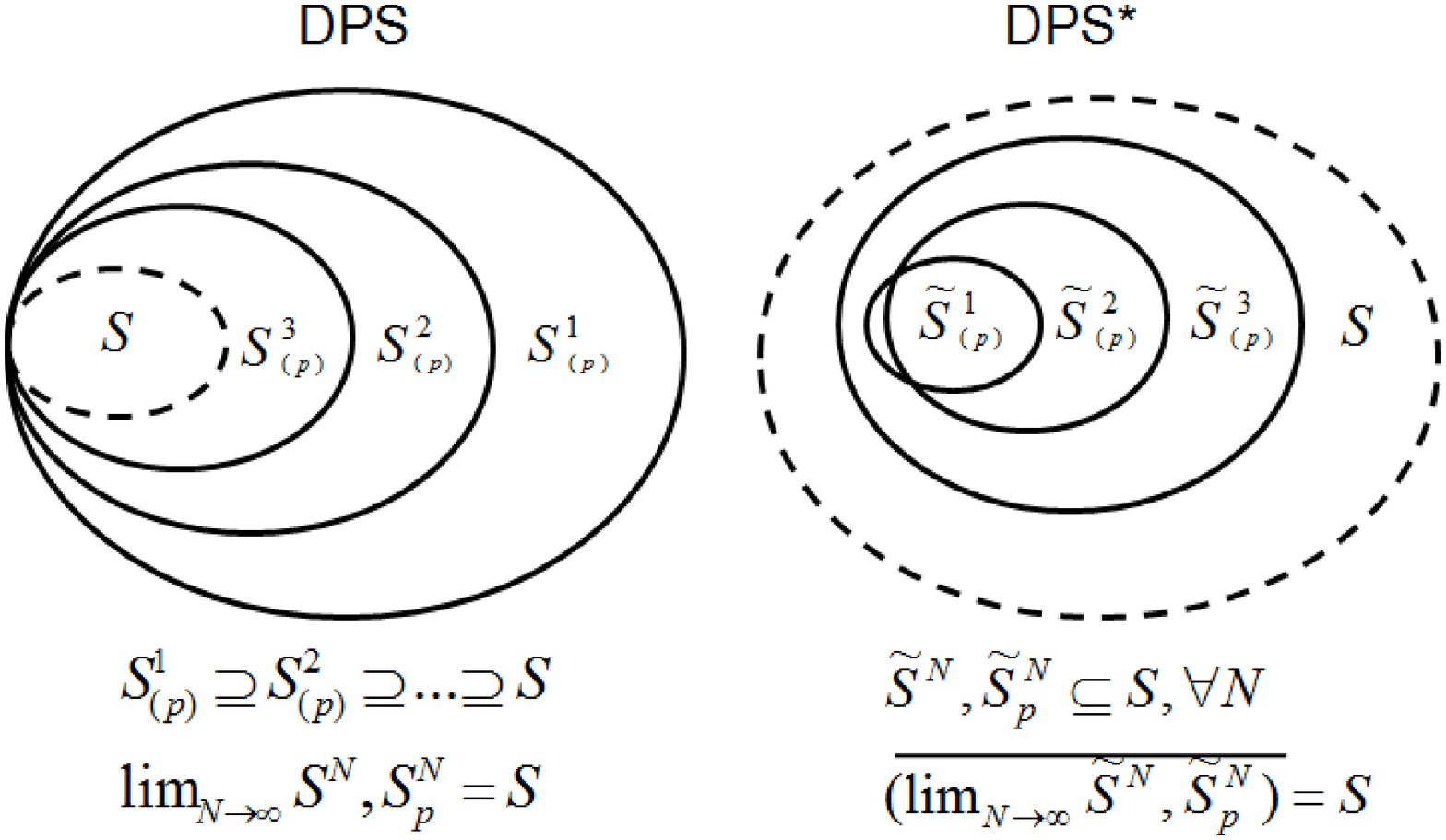}
  \caption{Two different ways of approximating the set $\S$ of separable operators (dashed line): from the outside (DPS criterion) or from the inside (DPS$^*$ criterion).}
  \label{DPS}
\end{figure}

We will now proceed to study the speed of convergence of DPS$^*$. First, notice that, as $\S\subset \S^N, \S_p^N$, 
equation (\ref{interior}) tells us that we can perturb any general separable state $\rho_{AB}$ to a state $\tilde{\rho}_{AB}$ 
inside ${\tilde \S}^N$ and ${\tilde \S}_p^N$. 
This observation, together with the techniques employed in \cite{pra}, allows us to derive the following proposition:

\begin{prop}
\label{distancias}
Denote by $\|\bullet\|_p$ the $p$-norm, i.e., $\|Z\|_p=(\tr(\sqrt{ZZ^\dagger}^p))^{1/p}$, and let $\rho \in \S$ be a separable state. 
Then, there exists a normalized state $\sigma\in\tilde{\S}^N$ such that:

\begin{eqnarray*}
\|\rho-\sigma\|_1\leq\frac{2(d-1)}{N+d}, \ {\rm and } \ \|\rho-\sigma\|_\infty\leq\frac{d-1}{N+d}.
\end{eqnarray*}
\noindent Analogously, there exists a normalized state $\sigma_p\in\tilde{\S}^N_p$ such that
\begin{eqnarray*}
\|\rho-\sigma_p\|_1\leq\frac{2(d-1)\epsilon_N}{d},\ {\rm and} \ \|\rho-\sigma_p\|_\infty\leq\frac{(d-1)\epsilon_N}{d}.
\end{eqnarray*}

\end{prop}

\noindent It is worth noting that all the above bounds cannot be improved, since they correspond 
to the exact distances of $\rho_{AB}$ to the sets $\tilde{\S}^N$ or $\tilde{\S}^N_p$ whenever $\rho_{AB}$ is a pure product state.

Proposition \ref{distancias} allows us to study the efficiency of the criterion based on the sequence of sets $(\tilde{\S}^N)$ as opposed to the one based on $(\tilde{\S}^N_p)$ when applied to solve the weak membership problem of separability (WMEM($\S$)) \cite{G03}, \cite{sevag}, i.e., the problem of determining whether a state $\rho_{AB}$ is separable or not up to some precision $\delta$. Since our bounds on the speed of convergence of DPS$^*$ have the same scaling as those derived for the DPS criterion in \cite{pra}, our conclusions on the complexity of the former cannot but be the same, namely:

\begin{enumerate}
\item
From the point of view of time complexity, it is always preferable to consider the sets $\tilde{\S}_p^N$ rather than $\tilde{\S}^N$. Whereas the time complexity of the latter scales with $\delta$ as $O\left(d_A^6(k_1/\delta)^{6d_B}\right)$, the number of operations required under the PPT constraint scales as $O\left(d_A^6(k_2/\delta)^{4d_B}\right)$.

\item
From the point of view of space complexity, sometimes it may be more convenient to search over the sets $\{\tilde{\S}^N\}$ 
rather than over $\{\tilde{\S}_p^N\}$.

\item
In any case, due to its fast proven convergence as compared to other methods \cite{I07}, 
the DPS$^*$ criterion is one of the most efficient algorithms for entanglement detection. Because of the polynomial dependence on $d_A$ of its bounds on time or space complexity, DPS$^*$ is specially useful to attack problems where the dimension of one of the systems is much bigger than the other one's.


\end{enumerate}










We will now illustrate the power and versatility of this algorithm by showing how to use it to obtain approximate solutions for two important problems that appear frequently in quantum information:

\noindent\emph{a) State Estimation Problems}

In a generic pure state estimation scenario, a source randomly chooses a pure state $\Psi_i$ out of a probability distribution $p_i$ and then encodes it into a state $\Psi_i'$ to which we are given full access. The state estimation problem consists on finding the optimal measure-and-prepare strategy that allows us to reconstruct the original state $\Psi_i$ with high fidelity.
In \cite{pasado}, it is explained how to map state estimation (SE) problem problem into a linear optimization over the set $\S$ of separable operators, 
via the relation
\begin{equation}
\hspace*{-1cm} F=\max\{\tr(\rho_{AB}\Lambda_{AB}):\Lambda_{AB}\in \S,\Lambda_A=\id\},
\label{funda}
\end{equation}

\normalsize

\noindent where $F$ is the optimal average fidelity and $\rho_{AB}=\sum_i p_i \Psi'_i\otimes\Psi_i$ is given by the particular SE problem. 
In \cite{pasado} it is also shown that any separable decomposition $\Lambda_{AB}=\sum_x M_x\otimes \phi_x$ of an operator $\Lambda_{AB}$ satisfying the above conditions can be interpreted as a measure-and-prepare strategy consisting of applying the POVM $\{M_x\}_x$ and preparing the state $\phi_x$ depending on the outcome $x$.

Now, consider the sequence of optimization problems:
\begin{eqnarray}
\hspace*{-1cm}& &\tilde{F}^N\equiv\max\{\tr(\rho_{AB}\tilde{\Lambda}_{AB}):\tilde{\Lambda}_{AB}\in
\tilde{\S}^N,\tilde{\Lambda}_A={\mathbb I}\},\nonumber\\
\hspace*{-1cm}& &\tilde{F}_p^N\equiv\max\{\tr(\rho_{AB}\tilde{\Lambda}_{AB}):\tilde{\Lambda}_{AB}\in
\tilde{\S}^N_p,\tilde{\Lambda}_A={\mathbb I}\}.
\label{ideacentral}
\end{eqnarray}

\normalsize

These optimizations can also be cast as a semidefinite program. 
Therefore, they can be efficiently computed as long as the index $N$ is not very high (note nevertheless, that, in general, if we fix $N$ for increasing $d_B$, we will end up with a very bad approximation). 
Also, because we are optimizing over particular regions of $\S$, in general we will get a suboptimal result, i.e., $\tilde{F}^N,\tilde{F}^N_p\leq F$ for all $N$. 
However, it is clear that $\lim_{N\to\infty}\tilde{F}^N,\tilde{F}^N_p=F$. 

As we have already mentioned, we can write down a separable decomposition for the operator $\tilde{\Lambda}_{AB}$ output by the computer, via
Eq.(\ref{eq: separable decomposition}) in case $\tilde{\Lambda}_{AB}\in \tilde{\S}^N$ or by means of a similar formula \cite{pra} in case $\tilde{\Lambda}_{AB}\in\tilde{\S}_p^N$.
This separable decomposition can be subsequently interpreted as a measure-and-prepare strategy; 
for any SE problem, the DPS$^*$ method can thus provide us with a sequence of state estimation strategies that converge asymptotically to the optimal one.

It would be natural to wonder how fast this convergence is, i.e., how far $\tilde{F}^N$ is from $F$ for finite $N$. 
Following the lines of \cite{pra}, we arrive at a sequence of upper and lower bounds on $F$ given by
\begin{eqnarray}
& &\tilde{F}^N\leq F\leq\tilde{F}^N+\frac{d}{N}\left(\tilde{F}^N-\frac{1}{d}\right),\nonumber\\
& &\tilde{F}_p^N\leq F\leq\tilde{F}^N_p+\frac{\epsilon_N}{1-\epsilon_N}\left(\tilde{F}^N_p-\frac{1}{d}\right).
\end{eqnarray}

To get an idea of the efficiency of these algorithms, we refer the reader to \cite{pra}, since the lower bounds on the maximal fidelity of the state estimation problems considered in that paper actually correspond to the values computed through DPS$^*$.

\noindent\emph{b) Computation of the robustness of entanglement}

Let $\rho \in {\cal B}(\H )$ be a quantum state. We will define its  \emph{(separable) robustness of entanglement} \cite{vidal} $R(\rho)$ as
\begin{eqnarray}
R(\rho )=\min\{\tr (\sigma) :\sigma \in \S, \ \rho +\sigma \in \S\}.
\end{eqnarray}

Used in combination with the DPS criterion, the DPS$^*$ criterion allows us to determine 
the robustness of entanglement of any quantum state $\rho_{AB}$ up to arbitrary precision. Indeed, note that, optimizing over $\S^N (\S_p^N)$ or ${\tilde \S}^N ({\tilde \S}_p^N)$ instead of $\S$ in the previous definition, 
we would obtain lower and upper bounds on $R(\rho)$, respectively. 
And, of course, both optimizations can be performed using semidefinite programming.

In \cite{datta}, the authors introduced a class of uniparametric families of $n$-partite quantum states. 
Given a unitary operator $V$ acting over $n-1$ qubits, the corresponding family of states is defined as
\begin{eqnarray*}
\hspace*{0.1cm}\rho^V_\alpha\equiv \frac{1}{2^n}\id_1\otimes \id_{23...n}+\frac{1}{2^n}\alpha\ket{0}\bra{1}\otimes V +\frac{1}{2^n}\alpha\ket{1}\bra{0}\otimes V^\dagger.
\end{eqnarray*}

The separability properties of these states are very important, for if one could prove that any such state $\rho^V_\alpha$ is multiseparable for all $\alpha\leq 1/\mbox{poly}(n)$, then one would have an example of a quantum computation that supersedes any classical algorithm but nevertheless does not require entanglement \cite{datta}. For the case $n=3$, the authors showed that, for any unitary $V$, the state $\rho^V_\alpha$ is always separable with respect to the partition $1|23$. Moreover, their numerical tests suggested that $\rho^V_\alpha$ is PPT with respect to the partition $12|3$ for all $\alpha\leq 1/2$. Nevertheless, they conjectured that, in some cases, the state should remain entangled for lower values of $\alpha$. 
The DPS$^*$ criterion presented above strongly suggests that this is not the case. 

We generated 1000 random unitaries $\{V_i\}_i\subset SU(4)$ according to the Haar measure and applied the DPS$^*$ criterion to derive upper bounds on the robustness of entanglement of the corresponding states $\{\rho^{V_i}_{0.5}\}_i$ by considering extensions over the last qubit. We used the MATLAB package \emph{YALMIP} \cite{yalmip} in combination with \emph{SeDuMi} \cite{sedumi} to perform the numerical calculations. After an appropriate optimization over the set $\tilde{\S}^{3}$, all the corresponding upper bounds turned out to be zero, thus proving the separability of the previous sample of states.

\begin{figure}
  \centering
  \includegraphics[width=9 cm]{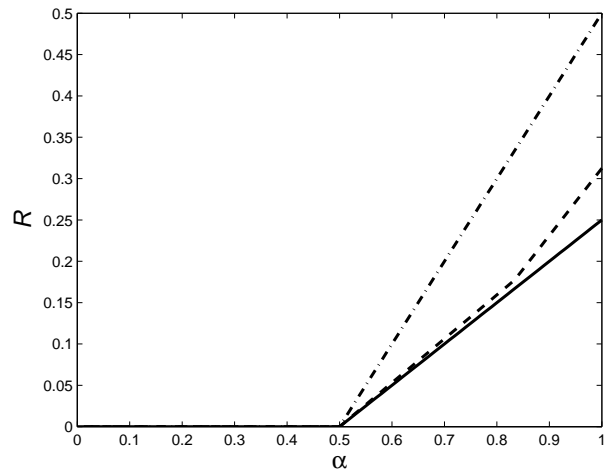}
  \caption{Upper (DPS$^*$) and lower (DPS) bounds on the entanglement robustness $R$ of $\rho^U_\alpha$ as a function of $\alpha$. The optimizations over the sets $\tilde{\S}^3$, $\tilde{\S}^{15}$ and $\S^3_p$ correspond to the dashed-dotted, dashed and solid lines, respectively.}
  \label{robustness}
\end{figure}

In order to discard statistical effects, we considered the unitary operator $U=2P_{\mbox{sym}}^2-\id_2\otimes\id_2$, whose associated states $\rho_\alpha^U$ appear to have the greatest negativity of the whole family for fixed $\alpha>0.5$ \footnote{Animesh Datta, private communication.}. Fig. \ref{robustness} shows upper (DPS$^*$) and lower (DPS) bounds on the robustness of entanglement for different values of $\alpha$. This time, we optimized over the sets $\tilde{\S}^{3}$, $\tilde{\S}^{15}$ and $\S^3_p$, respectively. It is clear that for all values of $\alpha$ below 0.5, the state $\rho^U_\alpha$ is separable.


In conclusion, in this letter we have introduced a new criterion for separability detection. 
This criterion has been inspired by and it is in a sense complementary to the one conceived by Doherty et al. \cite{doherty2}. 
Whereas the latter one aims at approximating the set of separable states from the outside, our new method works from the inside, 
i.e., by defining a sequence of sets of states contained in $\S$. From \cite{pra}, it follows that the method can be easily extended to deal with multiseparability problems.

This new criterion works basically by taking the states defined by Doherty et al. and applying a perturbation to make them separable. 
We believe, however, that the size of this perturbation is larger than required: in the PPT case, at least for small dimensions of Alice's system, a smaller transformation (in some cases the identity) should be enough to guarantee the separability of the output state. 
Note that the bounds on the distance to arbitrary separable states are actually independent of $d_A$. 
If future research found an optimal linear map to turn Doherty et al.'s states into separable states that took into account the dimensionalities of both systems, the speed of convergence of the resulting improved DPS$^*$ criterion would be much faster.

Nevertheless, a wide range of applications follow from our method in its present state. We have seen that the method alone can be used to determine the best experimental setup in quantum tomography protocols, 
and we have also shown how the combination of both the DPS and DPS$^*$ methods allows to solve with arbitrary precision computationally hard problems, 
like the calculation of the entanglement robustness. Since some important problems in Complexity Theory like PARTITION or CLIQUE can be reduced to the separability problem \cite{G03,sevag}, it is not unrealistic to expect that, with time, our method will find new applications outside the scope of Quantum Information Science.

The authors thank Animesh Datta and Fernando G. S. L. Brand\~{a}o for useful discussions. This work is part of the EPSRC QIP-IRC and is supported by EPSRC grant EP C546237 1, the Royal Society, the EU Integrated Project QAP and an Institute for Mathematical Sciences postdoc fellowship.

\end{document}